\documentclass{ifacconf}

\usepackage{booktabs} 
\usepackage{amsmath}
\usepackage{amssymb}
\usepackage{amsbsy}
\usepackage{algorithm}               
\usepackage{algpseudocode}
\usepackage{graphicx}  
\usepackage{psfrag}
\usepackage{diagbox}
\usepackage{bm}
\usepackage{bbm}                                                       
\allowdisplaybreaks

\usepackage{bm}
\usepackage{bbm}
    
\RequirePackage{epsfig}
\RequirePackage{psfrag}
\RequirePackage{pstool}
\RequirePackage{epstopdf}
\RequirePackage{graphicx}
\usepackage{bbm}
\usepackage{subfigure}
\RequirePackage{mathrsfs}

\RequirePackage{amsmath}
\RequirePackage{amssymb}

\RequirePackage{amsbsy}
\RequirePackage{amsfonts}
\RequirePackage{mathrsfs}
\usepackage{algorithm}   
\usepackage{algpseudocode}

\usepackage{psfrag} 
\usepackage{amsmath}

\newtheorem{remark}{Remark}

\usepackage{algorithm}   
\usepackage{algpseudocode}

\DeclareMathOperator*{\argmin}{arg\,min}

\usepackage{enumitem}
\usepackage{tabularx,booktabs,ragged2e}

\usepackage{graphicx}      
\usepackage{natbib}        
\begin{document}
\begin{frontmatter}

\title{On Optimal Synchronization of Diffusively Coupled Heterogeneous Van der Pol Oscillators \thanksref{footnoteinfo}}

\thanks[footnoteinfo]{Partial financial support by the German Research Foundation (DFG) through the Research Training Group "Biological Clocks on Multiple Time Scales" is gratefully acknowledged.}

\author[First]{Tabea Trummel}
\author[First]{Zonglin Liu}
\author[First]{Olaf Stursberg}

\address[First]{Control and System Theory, Dept. of Electrical Engineering and Computer Science, University of Kassel, Germany (e-mail: \{t.trummel, z.liu, stursberg\}@uni-kassel.de).}

\begin{abstract}                
This paper proposes a novel method to achieve and preserve synchronization for a set of connected heterogeneous Van der Pol oscillators. Unlike the state-of-the-art synchronization methods, in which  a large coupling gain is applied to couple any pair of connected oscillators, the proposed method first casts the whole synchronization process into two phases. The first one considers the period from the beginning to the first instant of synchronization, while the second phase covers the following time in which synchronization must be preserved. It is shown that a large  coupling gain is adopted for the first phase, while the averaged coupling gain  to preserve the synchronization in the second phase can be  reduced significantly by using an offline optimized coupling law. Efficiency and performance of this method are confirmed by a set of numerical tests with different graphs  and  system dynamics.
\end{abstract}


\end{frontmatter}

\section{Introduction} \label{INT}

The research on applying control and system theory to the study of biological and biochemical systems has attracted large interest in the last two decades. Among others, the investigation of  nonlinear  oscillatory networks which stem from  oscillatory rhythms of  humans or insects,  such as the sleep-wake cycle, has become a relevant question in this domain, and leads to interesting questions also from a system-theoretic perspective. A typically relevant question is, which type of networked dynamic model with oscillatory behavior is relevant to describe and explain biological effects.

A popular  model used often to describe the oscillations of single biological components is the \emph{Van der Pol oscillator}, see \cite{linkens1979theoretical, tegnitsap2021modeling}.  (Of course, this type of oscillator has attracted much attention also for technical applications.) When using this model for considering sets of coupled biological components (as very common  in biochemical systems), the aspect of synchronization of Van der Pol oscillators becomes relevant, see the overview by \cite{wang2022cell}. Typically, for a set of (almost) identical biological components  modeled by Van der Pol oscillators, the work by  \cite{rand1980bifurcation, banning2011dynamics, low2006coupled}  has shown how to achieve synchronization by taking into account the coupling topology represented by graphs. Later work considered that often coupled biological components are not identical, leading to research on synchronization of heterogeneous  Van der Pol oscillators.
A recent study of this problem by  \cite{lee2018heterogeneous} has revealed, that in order to achieve synchronization, a diffusive coupling law with large coupling gain (also called \textit{strong coupling})  should be applied. It is shown that the large coupling gain can suppress the heterogeneity and thus achieve the synchronization. In succeeding work by  \cite{lee2021design}, this  concept of strong coupling is further elaborated on to synchronize also other types of heterogeneous  oscillators, such as more general ones of \textit{Li\'{e}nard} type.

Strong coupling gains mean, however, in most cases for a given set of coupled oscillators that strong interaction require high consumption  of energy or other resources (not only but also for biological systems). Especially if the synchronization must be maintained for longer periods of time, high  consumption of resources makes it unrealistic so stay in this mode of operation. 
Thus,  the work in  \cite{fardad2012optimal} has proposed an optimal coupling law to reduce the peak value of the gains, but this strategy was only considered and formulated for the case of linear oscillators.  The optimal gain in that work was determined by solving a semi-definite programming (SDP) problem involving  linear matrix inequalities (LMI), while the  optimized gain turned out to be  a compromise of the synchronization error, the largest coupling gain, and the number of  coupling edges. The work by  \cite{mosebach2015lqr}, which is also limited to linear oscillators, computed the optimal coupling gain for  synchronization by using the technique of linear quadratic regulator. Note that the optimal gains obtained in that work are tailored to ensure a certain  synchronization rate, while the principle can  be well extended to reduce the coupling gains. 

It is worth to emphasize that a considerable gap exists between linear  oscillators and the considered type of Van der Pol oscillators. Typically, after synchronization   the linear oscillators will keep being synchronized  by deactivating the diffusive coupling law, while the synchronization of the latter type of oscillators will typically stop. This difference also motivates the necessity to reduce the coupling gains by preserving the synchronization for Van der Pol oscillators -- but this aspect has rarely been studied before.
According to the authors' knowledge, the only related work is that by   \cite{lee2021design}, who mentioned that it is unnecessary to constantly use a  large coupling gain to preserve synchronization, but the work does not propose an effective method  to reduce the gain.

Inspired by the work  in \cite{ramos2006piecewise}, where nonlinear oscillators are piecewise linearized in order to compute a limit cycle, the paper on hand proposes an effective method to  reduce the coupling gains while preserving the synchronization of heterogeneous Van der Pol oscillators. The proposed method casts the whole synchronization process into two phases, in which a large and constant coupling gain is applied in the first phase to achieve the synchronization, while time-varying and edge-specific coupling gains with much smaller average values are applied to preserve the synchronization in the second phase.

In the next section, the scheme to synchronize all oscillators to a common limit cycle by using a  large coupling gain is briefly reviewed. Section 3 describes how local Van der Pol oscillators are piecewise linearized along the common limit cycle, in order to determine the optimal coupling gains for preserving synchronization. This method is then tested for different settings in  Sec. 4, together with a discussion on how to improve the method further to enhance system performance, and effects of the system topology are discussed. The work is concluded in Sec. 5 together with an outlook on possible future work.

\section{Problem Description} \label{PRE}

Consider a set $\mathcal{N}=\{1,\ldots, n\}$ of heterogeneous Van der Pol oscillators, each of which has a local dynamics:
\begin{align} \label{eq:vdp}
\begin{bmatrix} \dot{x}_{i,1}(t) \\ \dot{x}_{i,2}(t) \end{bmatrix} = \begin{bmatrix} x_{i,2}(t) \\ -x_{i,1}(t) + \mu_i (1-x_{i,1}(t)^2) x_{i,2}(t) \end{bmatrix}
\end{align}
with time $t \in \mathbb{R}^{\ge 0}$ and a specific parameter $\mu_i > 0$ (leading to the notion of \emph{heterogeneous oscillators}). It is well-known for these oscillators that for any given initial state  $x_i(0) \ne 0$ of each oscillator $i \in \mathcal{N}$, the local state  $x_i(t) = [x_{i,1}(t),  x_{i,2}(t)]^\top$ converges to a  limit cycle for $t \to \infty$, and the course of this  limit cycle is determined by $\mu_i$. Thus, for the considered group of heterogeneous  Van der Pol oscillators with different  $\mu_i$, these will eventually converge to their own and specific limit cycle, thus the group does not show synchronization.

\subsection{Diffusive Coupling Law} \label{diffusive}

Now consider that the set of oscillators is coupled as represented by a  directed and strongly connected graph $\mathcal{G} = (\mathcal{N},\mathcal{E})$, in which $\mathcal{N}$ denotes the set of nodes (vertices) and $\mathcal{E}$ the set of  directed edges. In literature, a common way to  synchronize the heterogeneous oscillators is to couple the local dynamics by a so called  \textit{diffusive coupling law}, in which  an external input signal $u_{i}(t) \in \mathbb{R}^{2}$ is applied to the local oscillator $i \in \mathcal{N}$ according to:
\begin{align}\label{eq:local_static_coupling}
u_{i}(t) := k_c \sum\limits_{j \in \mathcal{N}_i} (x_j(t) - x_i(t)).
\end{align}
The scalar $k_c \ge 0$ represents the coupling gain between  two connected nodes. Let the set $\mathcal{N}_i$  denote the indices of neighboured nodes of $i$ in the sense that a directed edge  $e_{ij} \in \mathcal{E}$  from $j \in \mathcal{N}_i$ to $i$ exists. By using
$f_i(x_i(t))$ as a short form of the right-hand side of \eqref{eq:vdp},  the  local dynamics of any  $i \in \mathcal{N}$ under the effect of the coupling law \eqref{eq:local_static_coupling} is given by:
\begin{align}\label{eq:local_coupledynamic}
 \dot{x}_i(t) = f_i(x_i(t)) + u_{i}(t). 
\end{align}
By  collecting all local states $x_i(t)$ in one global state vector $x(t) = [x^\top_1(t) \ldots x^\top_n(t)]^\top$ and by using
$f(x(t))$ to denote the collection of local dynamics $[f^\top_1(x_1(t)) \ldots f^\top_n(x_n(t))]^\top$ of all oscillators, the overall coupled  dynamics follows to:
\begin{align} \label{eq:global_static_coupling}
\dot{x}(t) = f(x(t)) - k_c (\mathcal{L} \otimes I_2) x(t).
\end{align}
In here, $\otimes$ denotes the Kronecker product of two matrices, $I_{2}$ denotes an $2 \times 2$ identity matrix, and
$\mathcal{L}$ is the Laplacian matrix of the graph $\mathcal{G}$ (more details on  Laplacian matrices of graphs are defined e.g. in \cite{bullo2019lectures}).

\subsection{Synchronization to Limit Cycles of the Blended \\ Dynamics} \label{blend}

\begin{figure}[!b]
\centering
\includegraphics[scale=0.37]{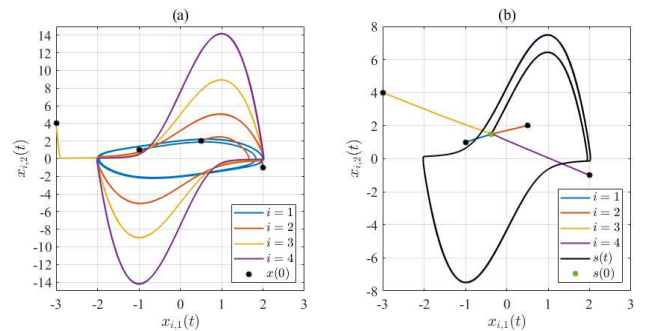}
\caption{(a) The limit cycle of a set of heterogeneous Van der Pol oscillators. (b) The set of oscillators synchronizes to the  limit cycle of  \eqref{eq:blended_dynamics} by using the diffusive coupling law \eqref{eq:local_static_coupling} with $k_c = 200$.}
\label{fig:limit_cycles}
\end{figure}

Based on \eqref{eq:global_static_coupling}, it is shown in \cite{lee2018heterogeneous} and \cite{lee2021design}, that for any synchronization accuracy $\epsilon >0$, there always exists a threshold $\kappa$ of the coupling gain $k_c$, such that:
 \begin{align} \label{eq:converge_blended}
\limsup\limits_{t \rightarrow \infty}||x_i(t) - s(t)|| \le \epsilon, \,\, \forall i \in \mathcal{N}
\end{align}
applies for all $k_c \ge \kappa$. Here,  $s(t) \in \mathbb{R}^{2}$ is the solution to the averaged dynamics of all oscillators (also called \textit{the blended dynamics}):
 \begin{align} \label{eq:blended_dynamics}
\dot{s}(t) = \frac{1}{n} \sum\limits_{i \in \mathcal{N}} f_i( s(t)),\,\, s(0) = \frac{1}{n} \sum\limits_{i \in \mathcal{N}} x_i(0).
\end{align}
It is further shown in the previously cited work, that the  blended dynamics \eqref{eq:blended_dynamics} also represents a Van der Pol oscillator (due to $\frac{1}{n}\sum\limits_{i \in \mathcal{N}} \mu_i>0$),  i.e., a limit cycle  also exists for $s(t)$. As a result,  the local states $x_i(t)$ of all oscillators $i \in \mathcal{N}$ are ensured to be  synchronized to the limit cycle of  \eqref{eq:blended_dynamics} with a desired accuracy  $\epsilon$, as long as not all  initial states   $x_i(0)$ are zero, see Fig.~\ref{fig:limit_cycles}.

According to this result,  the  considered synchronization task for a set of heterogeneous  Van der Pol oscillators can be achieved by adopting the diffusive coupling law  \eqref{eq:global_static_coupling} and using a  coupling gain $k_c$ that is large enough. However, a larger $k_c$  also implies a stronger interaction between two connected oscillators by generating the  external input $u_{i}(t)$  according to  \eqref{eq:local_static_coupling}. If such an interaction also results in  costs (e.g., if the local oscillators represent the dynamics of different cells, and the cell-to-cell  interaction consumes a certain amount of nucleotides or energy), an unnecessarily strong interaction should thus be avoided over longer periods of time. 

Furthermore, after the  $x_i(t)$  are synchronized to the limit cycle of  \eqref{eq:blended_dynamics}, the diffusive coupling law must be further applied in order to  preserve synchronization. Otherwise, the  local $x_i(t)$  will immediately start to converge back to their local limit cycles once  \eqref{eq:global_static_coupling} is deactivated (implying that $k_c=0$). 
As the latter phase often represents a much longer time interval than the first one, the question arises of how to reduce the  interaction between the oscillators, while preserving synchronization in the sense that an upper bound of the distancees between $x_i(t)$ and $s(t)$ must not be exceeded.

With respect to this problem, a novel optimal diffusive coupling law is next proposed, which allows to adopt different coupling gains for different edges of the graph, as well as varying the gains over time. It will be shown that by using the considered scheme, the coupling gains of all edges can be reduced to   a much lower level compared to the threshold $\kappa$ for most of the time, while  synchronization is  preserved with an acceptably small deviation between the local states.

\section{Optimal Coupling Gains to Preserve Synchronization} \label{OPT}

As mentioned before, for  \eqref{eq:converge_blended} and  \eqref{eq:blended_dynamics}, it is known that by using a large enough  gain $k_c$ all  oscillators $i \in \mathcal{N}$ can be synchronized from different initial states to the limit cycle of  \eqref{eq:blended_dynamics} with an accuracy of $\epsilon$. The design of optimal  coupling gains to preserve the synchronization begins from partitioning the  limit cycle of  \eqref{eq:blended_dynamics}: 
The  period $T$ of the  limit cycle of  \eqref{eq:blended_dynamics}, is divided into $f$ intervals of lengths  $\Delta t > 0$, i.e. $T=f \cdot \Delta t$. By selecting $t_0$ as starting time of a period, the sequence of sampling times   $\phi_t :=(t_0, t_1, \ldots,t_{f-1})$ results with $t_l= t_0 + l \cdot \Delta t$ for all $l \in \{0,\ldots, f-1\}$. For an arbitrary $s_0\in \mathbb{R}^{2}$ on the limit cycle of  \eqref{eq:blended_dynamics}, a state sequence $\phi_s:=(s_0, s_1, \ldots,s_{f-1})$ with $s_0=s(t_0)$ is obtained and records the states on  the limit cycle for $t_l$,  $l \in \{0,\ldots, f-1\}$.  
The local dynamics \eqref{eq:vdp} is linearized for all $i \in \mathcal{N}$ along the  state sequence $\phi_s$. By linearizing the local dynamics  of any $i  \in \mathcal{N}$ at any state $s_l$ in $\phi_s$, the following linearized dynamics is obtained:
\begin{align}\label{eq:taylor}
 \dot{x}_i(t)  = A_{i,l} x_i(t)  + b_{i,l}, 
\end{align}
with  $A_{i,l} := \begin{bmatrix} \frac{\partial f_i(x_i)}{\partial x_i} \end{bmatrix}  \Big{|}_{x_i = s_l}$ and $b_{i,l}:=  f_i(s_l) - A_{i,l} s_l$. 

The diffusive coupling law   \eqref{eq:local_static_coupling} is also extended to allow different edges $e_{ij} \in \mathcal{E}$ to use different coupling gains instead of using $k_c$ for all edges. 
With $K_{ij}(t) = \text{diag}(k_{ij,1}(t),  k_{ij,2}(t)) \in \mathbb{R}^{2 \times 2}$, the new diffusive coupling law is written as:
\begin{align}\label{eq:local_edgewise_coupling}
u_{i}(t) := \sum\limits_{j \in \mathcal{N}_i} K_{ij}(t) (x_j(t) - x_i(t)).
\end{align}
Note that   $K_{ij}(t)$ must be  a diagonal matrix according to the rule of diffusive coupling, see  \cite{bullo2019lectures}, since the value of the resulting $u_{i}(t)$ in each dimension can only depend on the difference $x_j(t) - x_i(t)$ for the associated oscillator.
 By using $K_{ij, l}$ to denote the value of  $K_{ij}(t)$ in sampling time $t_l$,  $l   \in \{0,\ldots, f-1\}$, and  by assuming that $K_{ij}(t) := K_{ij, l}$ holds  until $t_{l+1}$, the  linearized dynamics of the $i$-th
 oscillator with the coupling law  \eqref{eq:local_edgewise_coupling} takes the form:
\begin{align}\label{eq:locallinearcoupled}
 \dot{x}_i(t)  = A_{i,l} x_i(t)  + b_{i,l} + \sum\limits_{j \in \mathcal{N}_i} K_{ij, l} (x_j(t) - x_i(t)).
\end{align}
For the global vector $x(t)$ containing all local states, 
define the matrices:
$A_l := \text{diag}([A_{1,l}, \ldots, A_{n,l}])$, $b_l =[ b^\top_{1,l},\ldots,  b^\top_{n,l}]^\top$, and $L_{K, l}$ satisfying:
\begin{align}\label{eq:laplace_matrix_w}
L_{K, l} = \begin{bmatrix} M_{11} & M_{12} & \dots & M_{1n} \\ M_{21} & M_{22} & \ddots & \vdots \\ \vdots & \ddots & \ddots &  \vdots \\ M_{n1} & \dots & \ddots & M_{nn} \end{bmatrix} 
\end{align}
with $M_{ii}:=  \sum\limits_{j \in \mathcal{N}_i} K_{ij, l}$ for all $i \in \{1,\ldots, n\}$, and  $M_{ij}:= - K_{ij, l}$ if $e_{ij} \in \mathcal{E}$  for all $i \ne j$, 
and $M_{ij}:=0_{2 \times 2}$ otherwise. 
The index $K$ in  $L_{K, l}$ is used to emphasize that the value of $L_{K, l}$ depends on the set of coupling gain matrices $\{K_{ij, l}\}$.
The global dynamics over all oscillators is then:
\begin{align}\label{eq:lin_global_dynamic_with_Lw}
\dot{x}(t) = (A_l - L_{K, l})  x(t) + b_l.
\end{align}

The determination of optimal gain matrices  $\{K_{ij, l}\}$ for any $t_l$, can be carried out bases on \eqref{eq:lin_global_dynamic_with_Lw}. 
First of all, a value function $V(x(t)): \mathbb{R}^{2n} \to  \mathbb{R}$ is defined which records the overall synchronization error between any pair of  oscillators (which are not necessarily connected since all oscillators are assumed to be synchronized to a common limit cycle).
By use of a positive semi-definite matrix:
\begin{align} \label{eq:matrixp}
P = \begin{bmatrix} n-1 & -1 & \dots & -1 \\ -1 & n-1 & \ddots & \vdots \\ \vdots & \ddots & \ddots & -1 \\ -1 & \dots & -1 & n-1 \end{bmatrix} \otimes I_2
\end{align}
the value function is selected to:
\begin{align}\label{eq:error}
V(x(t)) := \hspace{-3mm}\sum\limits_{i, j \in \mathcal{N}, \, i \ne j} \hspace{-1mm} \frac{1}{2} (x_j(t) - x_i(t))^2  =x^\top(t) P x(t) .
\end{align}
In order to ensure that the overall synchronization error  decreases along $t_l$, the derivative of $V(x(t))$ with respect to $t$  should be smaller than zero in each $t_l$, i.e. the following inequality must hold according to   \eqref{eq:lin_global_dynamic_with_Lw} and \eqref{eq:error}:
\begin{align}
\dot{V}(x(t)) & =  x(t)^\top((A_l - L_{K, l})^\top  P + P (A_l - L_{K, l})) x(t) \notag\\
& \qquad + b^\top_l   P x(t) + x(t)^\top  P b_l <0.  \label{eq:V_dot}  
\end{align} 
In case the synchronization error  $x_j(t) - x_i(t)$ for all $i, j \in \mathcal{N}, \, i \ne j$ is small (since the considered optimized coupling law is applied to preserve the synchronization), the sum of the last two terms in  \eqref{eq:V_dot} approximately equals to zero for all   $b_l$ obtained from  linearization. As a result, the  inequality:
\begin{align}
 x(t)^\top(A_l - L_{K, l})^\top  P + P (A_l - L_{K, l}) x(t) <0 \label{eq:V_dot2} 
\end{align} 
is considered, which  is an approximation to   \eqref{eq:V_dot}. 

Now, regarding to the linearized dynamics \eqref{eq:lin_global_dynamic_with_Lw} and the inequality   \eqref{eq:V_dot2}, the 
 following optimization problem is proposed to determine the   optimal gain matrices  $\{K_{ij, l}\}$ in each sampling time $t_l$:  
\begin{subequations}\label{eq:opt}
\begin{align}
& \{ K^*_{ij, l}\}:=\argmin_{\alpha_l, \beta_l, K_{ij, l} \, \forall \, e_{ij} \in \mathcal{E}} \alpha_l + \omega \cdot \beta_l \label{eq:opt_min}\\
\text{s.t.:} \quad &(A_l - L_{K, l})^\top  P + P (A_l - L_{K, l}) < \alpha_l  \cdot I_{2n} \label{eq:opt_a}\\
& K_{ij, l} - \beta_l  \cdot I_2 < 0, \, \, \forall e_{ij} \in \mathcal{E}  \label{eq:opt_b}\\
& K_{ij, l} \ge   0_{2\times2}, \, \, \forall e_{ij} \in \mathcal{E} \label{eq:opt_c}\\
& \beta_l \geq 0 \label{eq:opt_d}
\end{align}
\end{subequations}
where $\omega \ge 0$ is a weighting factor,   $I_{2}$ and $I_{2n}$ are identity matrices with corresponding dimension, while $\alpha_l$ and $\beta_l$ are two new  variables to be optimized. The constraint  \eqref{eq:opt_a} originates from the inequality    \eqref{eq:V_dot2}, since the latter inequality is satisfied if $(A_l - L_{K, l})^\top  P + P (A_l - L_{K, l})$ is a  negative definite matrix (and thus the largest eigenvalue is smaller than zero due to the symmetric form). However, the local $x_i(t)$ is expected to follow a certain limit cycle in the phase of preserving synchronization. Thus, the global vector $x(t)$ in this phase only takes values close to the limit cycle (rather than from the whole space $ \mathbb{R}^{2n}$. Requiring  $(A_l - L_{K, l})^\top  P + P (A_l - L_{K, l})$ to be a  negative definite matrix would be too conservative,  since this implies  that \eqref{eq:V_dot2} were satisfied for all $x(t) \in  \mathbb{R}^{2n}$), thus leading to infeasibility of the optimization problem. Hence, 
a new variable  $\alpha_l$ is defined in  \eqref{eq:opt_a} and minimized in   \eqref{eq:opt_min}. Note that the variable  $\alpha_l$   in  \eqref{eq:opt_a}  represents an upper bound to the largest eigenvalue of $(A_l - L_{K, l})^\top  P + P (A_l - L_{K, l})$ according to \cite{boyd1994linear}.  Minimizing   $\alpha_l$ consequently also implies to reduce the value of $\dot{V}(x(t))$ based on  \eqref{eq:V_dot} and  \eqref{eq:V_dot2}.
 
In addition to   $\alpha_l$, the variable $\beta_l  \geq 0$ is also minimized in    \eqref{eq:opt_min}, which establishes an upper bound for all coupling gains corresponding to the edges  $e_{ij} \in \mathcal{E}$ according to  \eqref{eq:opt_b}. The minimization of  a weighted combination of $\alpha_l$ and $\beta_l$ leads to coupling gain matrices $ \{ K^*_{ij, l}\}$ that represents the combined consideration of the synchronization performance and the coupling effort (in terms of the gains being applied). Clearly, a larger $\omega$ results in smaller gains and a larger synchronization error, and vice versa. This scheme provides the possibility to determine a compromise between the two objectives, instead of simply using a coupling gain that is large enough. Moreover, this scheme can be utilized to study how the graph topology and the  gains required for preserving  synchronization affect each other, e.g. to determine which edge (with associated gain) is most important to enhance the synchronization performance most significantly. Finally, the coupling gains must non-negative as formulated in   \eqref{eq:opt_c}, as is common requirement for diffusive coupling according to  \cite{bullo2019lectures}. 
 
 \begin{remark}
 Note that the problem  defined by \eqref{eq:opt_min} -  \eqref{eq:opt_d} is not an LMI-constrained problem, since the matrix $ K_{ij, l}$ has to follow a certain structure leading to  bilinear matrix inequalities (BMI). For the numerical examples in the next section, the BMI-constrained problem is solved by using the solver PENBMI, see \cite{kocvara2005penbmi}. Nevertheless, as the constraint \eqref{eq:opt_c} is the only hard constraint $K_{ij, l}$ (while the other constraints are relaxed by the slack variables $\alpha_l$ and $\beta_l$), feasible solutions always exist in this problem.
 \end{remark}

Now, by solving the problem  \eqref{eq:opt_min} -  \eqref{eq:opt_d} for each sampling time $t_l$,  $l   \in \{0,\ldots, f-1\}$, a sequence of optimal coupling gain matrices:
\begin{align}\label{eq:opt_Kij} 
\phi_{K^*}:= (\{ K^*_{ij, 0}\}, \{ K^*_{ij, 1}\}, \ldots, \{ K^*_{ij, f-1}\} )
\end{align}
over the period $T$ of the limit cycle of  \eqref{eq:blended_dynamics} are obtained. This sequence of coupling gain matrices is then repeatedly applied over each period, in order to preserve the synchronization of all oscillators, see Algo. 1.

 \begin{algorithm} [!t]
 \caption{Two-phase synchronization method for heterogeneous  Van der Pol oscillators}
\begin{algorithmic}[1]

 \State \textbf{Given:}  Dynamics of the local Van der Pol oscillators \eqref{eq:vdp} and the initial states $x_i(0)$ for all $i \in  \mathcal{N}$, the graph  $\mathcal{G} = ( \mathcal{N},\mathcal{E})$, the  sampling interval $\Delta t$, and the weighting factor $\omega$.

 \State \textbf{Offline:}
 
 \State Determine the blended dynamics  \eqref{eq:blended_dynamics} and the corresponding limit cycle;
 \State Record the period $T$ of the   limit cycle and divide it into $f$ intervals using $\Delta t$;
  \State  Determine the  state sequence $\phi_s$ at each sampling time along the  limit cycle;
  \State Linearize the local dynamics in all states of  $\phi_s$ , and determine the sequence  $\phi_{K^*}$ in \eqref{eq:opt_Kij}  by solving the problem  \eqref{eq:opt_min} -  \eqref{eq:opt_d}.
  
   \State \textbf{Online:}

 \State  \textbf{Phase one: Achieve synchronization}
 \State Adopt the diffusive coupling law   \eqref{eq:local_static_coupling}  with a sufficiently large coupling gain $k_c$, until all local states $x_i(t)$ are synchronized to the limit cycle of   \eqref{eq:blended_dynamics} with specified tolerance $\|s(t_l)-x_i(t_l)\|_2\leq \epsilon\ \forall\ t_l\in\phi_t$ and $\epsilon\in\mathbb{R}^{>0}$.
   
  \State \textbf{Phase two: Preserve synchronization}

 \State After the   local states are synchronized and the state $s_0$  is reached for the first time by all $i \in  \mathcal{N}$, switch the coupling law  from \eqref{eq:local_static_coupling}   to   \eqref{eq:local_edgewise_coupling}.  
  \State  Sequentially apply the coupling gain matrices in $\phi_{K^*}$,  each for the duration $\Delta t$. After the last entry $\{ K^*_{ij, f-1}\}$ of  $\phi_{K^*}$ has been applied for $\Delta t$, switch the coupling gains to the first entry $\{ K^*_{ij, 0}\}$ of  $\phi_{K^*}$ and repeat the procedure of phase two.

 \end{algorithmic}
\end{algorithm}

\section{Numerical Examples}

To evaluate the performance of Algo.~1, a set of  $n = 4$ heterogeneous  Van der Pol oscillators with parameters $\mu_1 = 0.5$, $\mu_2 = 3$, $\mu_3 = 6$ and $\mu_4 = 10$ are considered. For the case without coupling, the limit cycles of the oscillators as illustrated in Fig.~\ref{fig:limit_cycles}(a) occur.  

\begin{figure*}[!t]
\centering
\includegraphics[scale=0.42]{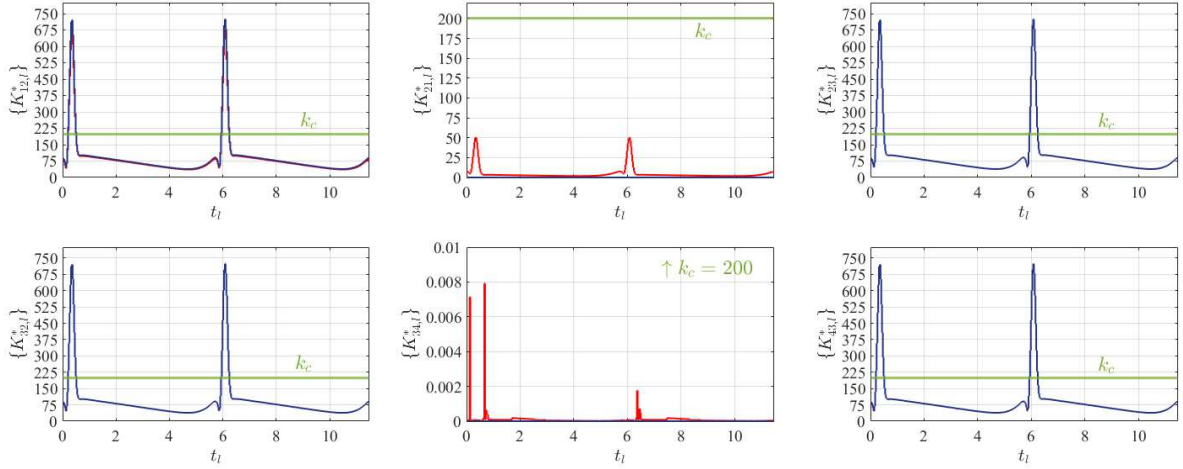}
\caption{Optimized  coupling gains $\{ K^*_{ij, l}\}$ for all 6 edges in the chain-shaped graph within one period {\color{blue} $T=11.48$}. Note that {\color{blue} most of} the two entries of the diagonal matrix $K^*_{ij, l}$ are taking similar values for the whole period, leading to almost identical trajectories marked in red and blue.}
\label{fig:chain_Kij}
\end{figure*}

\begin{figure}[!b]
\centering
\includegraphics[scale=0.39]{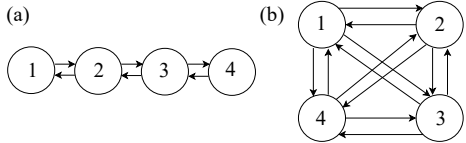}
\caption{Chain-shaped graph (a) and  fully connected graph (b).}
\label{fig:net_structures}
\end{figure}

In the first test, the oscillators are assumed to be connected according to a chain-shaped graph as shown in Fig.~ \ref{fig:net_structures}(a). In the offline  part of Algo.~1, the limit cycle of the blended dynamics \eqref{eq:blended_dynamics} (see Fig..~\ref{fig:limit_cycles}(b)) is first determined and the whole period {\color{blue} $T=11.48$} is sampled into $f = 400$ intervals. The optimal coupling gain matrices $\{ K^*_{ij, l}\}$ at each sampling time $t_l$,  $l \in \{0,\ldots, f-1\}$ are determined by solving the  problem  \eqref{eq:opt_min} - \eqref{eq:opt_d}, see the edgewise illustration within one period in Fig.~\ref{fig:chain_Kij}.

\begin{figure}[!b]
\centering
\includegraphics[scale=0.44]{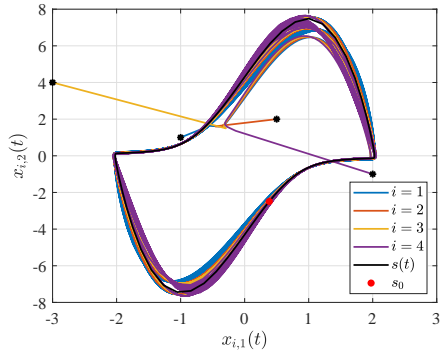}
\caption{Synchronization is achieved and preserved for 20 periods by using Algo.~1 (chain-shaped graph).}
\label{fig:chain_state_space}
\end{figure}

Clearly, compared to the coupling gain $k_c = 200$ to be used to achieve synchronization in the first phase of  Algo.~1, the optimized gains in $\{ K^*_{ij, l}\}$ are far below $k_c$ for almost all edges and for the most time of a period, with an average gain of {\color{blue} 90} over all edges (after averaging the gain of each edge over the period). For the online part of  Algo.~1, the oscillators are first synchronized by using the coupling law \eqref{eq:local_static_coupling} with given $k_c$, and then they switch to the coupling law \eqref{eq:local_edgewise_coupling} with the optimized coupling gains when the selected state $s_0$ is reached. The resulting trajectory of each oscillator by sequentially applying the optimized coupling gains in $\phi_{K^*}$ for 20 periods, i.e., for a total time of {\color{blue} $20 \cdot T = 229.6$}, are shown Fig.~ \ref{fig:chain_state_space}. 

\begin{figure}[!b]
\centering
\includegraphics[scale=0.44]{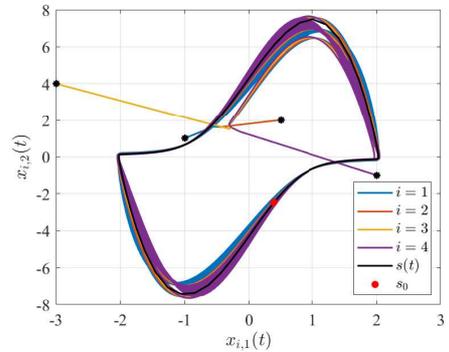}
\caption{By allowing the oscillators to temporarily switch back to the strong coupling gain in order to mitigate accumulated synchronization errors, the set of oscillators can preserve satisfiable synchronization even after $200$ periods.}
\label{fig:chain_state_space_restarts_10}
\end{figure}

Note that the local oscillators can be further synchronized (with acceptable deviation around the limit cycle of the  blended dynamics) for the whole time, although  much smaller coupling gains are used. Nevertheless, it can be observed, that the synchronization error accumulates by using the optimized coupling gains, and the maximal deviation is increasing over time (see the lower left part of the limit cycle). To address this problem, one can simply temporarily switch the coupling law back to \eqref{eq:local_static_coupling} with a large $k_c$, until they are re-synchronized. Afterwards, the optimal coupling gains can be further adapted, until a threshold of the maximal synchronization error is reached. This hybrid scheme, in general, only requires a large coupling gain be activated for a very short time, while smaller gains are used for the most of the time. An example of using this hybrid scheme for over $200$ periods of time is shown in Fig.~ \ref{fig:chain_state_space_restarts_10}.

\begin{figure}[!b]
\centering
\includegraphics[scale=0.44]{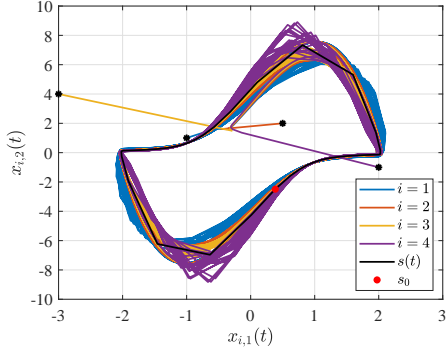}
\caption{Synchronization with $f = 100$ for 20 periods by using Algo.~1. A larger synchronization error than in Fig.~\ref{fig:chain_state_space} can be observed.}
\label{fig:chain_state_space_f_100}
\end{figure}

For the case that the period $T$ is divided into much less intervals, e.g., $f = 100$, implying that each optimized coupling gain $\{ K^*_{ij, l}\}$ must be hold constant four times longer than before, the  trajectories of all oscillators obtained from using Algo.~1 are shown in Fig.~\ref{fig:chain_state_space_f_100}. In general, a larger sampling time $\Delta t$ can result in worse synchronization performance. A possible solution to this problem is to adapt the weighting factor $\omega$ according to $\Delta t$ during the offline design phase.
 
\begin{figure}[!b]
\centering
\includegraphics[scale=0.44]{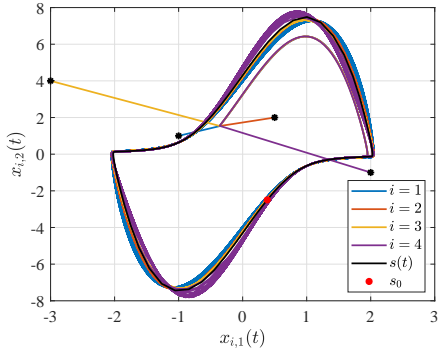}
\caption{Synchronization is achieved and preserved for 20 periods by using Algo.~1 (fully connected graph).}
\label{fig:fully_connected_state_space}
\end{figure}

\begin{figure}[!b]
\centering
\includegraphics[scale=0.40]{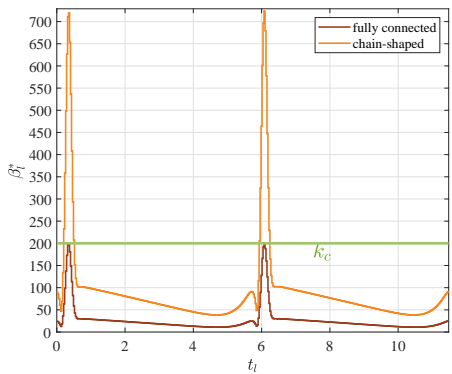}
\caption{The largest coupling gain over all edges for the chain-shaped graph, and the fully connected graph respectively, within one period.}
\label{fig:beta_chain_and_fully}
\end{figure}

For the same oscillators and same initial states, but for a full connection of the oscillators according to the graph in  Fig.~\ref{fig:limit_cycles}(b), the newly optimized coupling gains  $\{ K^*_{ij, l}\}$ can achieve a comparable synchronization performance as for the chain-shaped graph, see in Fig.~\ref{fig:fully_connected_state_space}. The averaged gains for the new variant turn out to be smaller, however, see the comparison  in Fig.~\ref{fig:beta_chain_and_fully}. This result (which is confirmed by observations for other graph structures) documents the effect that the {\color{blue} bigger the neighbourhood $\mathcal{N}_i$ of each oscillator in the graph is} the smaller is the required averaged coupling gain of each edge to preserve synchronization.

\section{Conclusions}

In this paper, an optimized diffusive coupling law to preserve the synchronization of a set of heterogeneous Van der Pol oscillators is proposed. Unlike the known approach of using a large and constant coupling gain between  any pair of connected oscillators to preserve the synchronization, the new method  determines a time-varying  gain for each edge in order to find the best compromise between the maximal gain value and the synchronization error. Effectiveness of this method is also confirmed in different simulations with significantly different oscillator parameterizations and different coupling topologies.  Future work aims at taking into account robustness of the optimized coupling laws for the case of uncertain components in the oscillator dynamics (which may also arise from identifying the oscillators from experimental data).

\bibliography{IFACWC_Bibliography}             

\end{document}